\documentclass{PoS}

\title{Practical methods for a direct calculation of $\Delta I=1/2$ $K$ to $\pi\pi$ Decay}

\ShortTitle{Methods for calculation of $K$ to $\pi\pi$ Decay}

\author{\speaker{Qi Liu}\\
        Department of Physics, Columbia University, New York, NY 10025, USA \\
        E-mail: \email{ql2142@columbia.edu}}

\author{RBC and UKQCD collaborations}

\abstract{
A direct calculation of the complex $\Delta I=1/2$ kaon decay amplitude is notoriously difficult because of the presence of disconnected graphs. Here we describe and demonstrate two practical methods to defeat this problem: the EigCG algorithm and the use of time-separated $\pi-\pi$ sources. With a fine tuned EigCG implementation for domain wall fermions, the calculation of light quark propagators is accelerated by a factor of 5-10 on a variety of lattices from small
($16^3\times32\times16$) to large ($32^3\times64\times32$). In addition, a substantial reduction in noise is achieved by separating each of the sources for the two pions in the time direction by 2-5 lattice spacings. These methods are combined in a calculation of $K$ to $\pi\pi$ threshold decay using a $24^3\times64\times16$ volume and $329$ MeV pions. These methods result in non-zero signals for both Re($A_0$) and Im($A_0$) from 138 gauge configurations.
}

\FullConference{XXIX International Symposium on Lattice Field Theory\\
		July 10-16, 2011\\
		Squaw Valley, Lake Tahoe, California}

\begin{document}

\section{Introduction}
To qualitatively understand the experiment phenomena of the $\Delta I=1/2$ enhancement rule and the direct CP violation in the neutral kaon decay process, a direct calculation of the $K^0\rightarrow\pi\pi$ weak matrix elements is needed. This includes the calculation of disconnected graphs and therefore requires very large statistics. Our previous systematic study of a full first principle calculation of  kaon decay on a $16^3\times32\times16$ volume lattice with Domain Wall Fermion (DWF)
showed promising results~\cite{Blum:2011pu} and encouraged us to try to approach physical kinematics on a larger volume. Our ultimate goal is to calculate the $\Delta I=1/2$ decay amplitude with all physical parameters just as we have done for the $\Delta I=3/2$ decay~\cite{Elaine2011}. Then we can compare the lattice results directly with experiment, which will provide us a deeper understanding of the $\Delta I=1/2$ rule and a check of the fundamental mechanism of CP
violation of the standard model.

While a full calculation with physical kinematics is still out of reach, we extended our previous calculation to a larger lattice with volume $24^3\times64\times16$, and decreased the pion mass from the previous 420 MeV to 330 MeV. A simple estimation of the volume effect and the number of CG iterations shows that this calculation for each configuration is 27 times more difficult than the previous one. Therefore, in this work, we concentrate on techniques to reduce the
difficulty of such a direct calculation. In the following, two techniques will be discussed in detail, first the EigCG Algorithm and then the method with time-separated $\pi-\pi$ sources. At the end, we will present our latest results for both $A_0$ and $A_2$ from the larger lattice with both techniques incorporated.

\section{Setup for the $K^0$ to $\pi\pi$ decay calculation}
The effective weak Hamiltonian for the $K^0$ to $\pi\pi$ decay including 2+1 flavors is
\begin{equation}
 H_{eff}=\frac{G_F}{\sqrt{2}}V_{ud}^*V_{us}
                      \sum_{i=1}^{10}[(z_i(\mu)+\tau y_i(\mu))] Q_i.
\label{eq:weak_eff}
\end{equation}
where $z_i$ and $y_i$ are the Wilson coefficients, $Q_i$ are the ten four-fermion operators. For more details about the effective weak Hamiltonian, the calculation of Wilson coefficients, and the definition of the four-fermion operators, see ref.~\cite{Buchalla:1995vs}. To obtain the decay amplitudes, we need to calculate the weak matrix element $<\pi\pi|Q_i|K^0>$ for each of the ten operators on the lattice, then convert them to the $\overline{\rm MS}$ scheme, and finally combine
with the Wilson coefficients which are also calculated in the $\overline{\rm MS}$ scheme. As we have describe this in detail in the Appendix A of \cite{Blum:2011pu}, the conversion from the lattice operators into the $\overline{\rm MS}$ scheme involves two steps. First we convert it into RI/MOM scheme, and then  convert the RI operators to the $\overline{\rm MS}$ scheme.

\begin{figure}[!tbh]
\begin{tabular}{cc}
\includegraphics[width=0.4\textwidth]{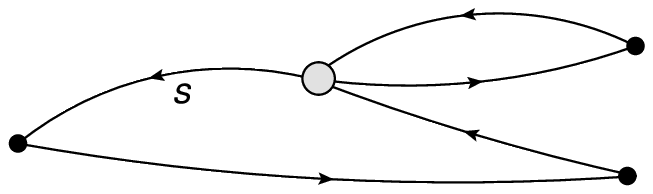} &
\includegraphics[width=0.4\textwidth]{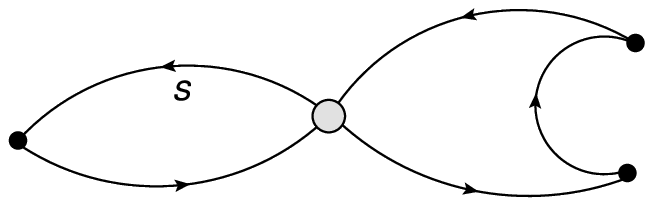} \\
Type 1 & Type 2 \\
\includegraphics[width=0.4\textwidth]{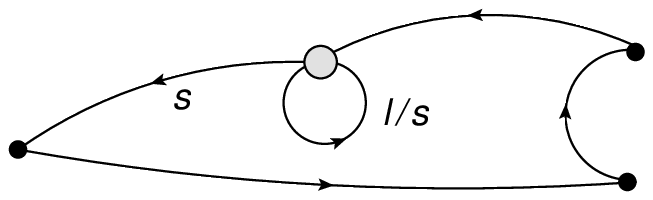} &
\includegraphics[width=0.4\textwidth]{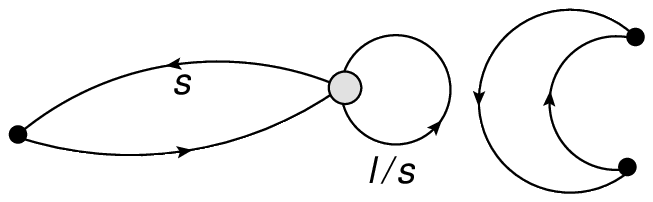} \\
Type 3 & Type4 \\
\end{tabular}
\caption{The four types of contractions that contributes to the calculation of $K^0$ to two pions decay. The graph circle stands for one of the four-fermion operators, the lines indicate the propagators(with addition label s meaning strange quark and otherwise light quark), and the black dot stands for the kaon or the pion with a $\gamma_5$ matrix insertion. The type 4 graph is the disconnected graph.}
\label{fig:contractions}
\end{figure}

The most important part of this work is to calculate the weak matrix elements on the lattice $Q_i^{lat}$. There are four types of contraction as shown in Figure\ref{fig:contractions}. The detailed structures of the different kinds of spin and color contractions for each type and related subtraction graphs are discussed in our previous work~\cite{Blum:2011pu}. 

We use a Coulomb Gauge fixed wall source and sink for the pions and kaons. Because of the presence of the disconnected graph (type 4), we are required to be able to put the sources on all possible time slices. Therefore, T (the dimension in the time direction) propagators with a wall source for both light and strange quarks are calculated. Even though computationally very expensive, it gives us the freedom to translate the position of the kaon, the operator and the two pions
simultaneously, thus effectively increasing the statistics from a given configuration. In addition to the Coulomb gauge wall source propagators, we also calculate T random wall source propagators to estimate the loop shown in the type 3 and 4 contractions by stochastic method.
In total, we need to solve $2T$ light quark propagators: on a typical T=64 lattice, this is equivalent to 1536 Dirac operator solves. At this point, it is clear to us that a good algorithm to speed up propagator calculation is crucial for such a calculation to be manageable. 

\section{The EigCG Algorithm}
There are two recently published algorithms for the calculation of propagators that could potentially provide a factor of 5-10 speed up. The first one is L\"uscher's inexact low modes deflation algorithm with the domain-decomposed subspaces that are based on the property called local coherence of the low modes~\cite{Luscher:2007se}. The second one is the EigCG algorithm by Stathopoulos and Orginos~\cite{Stathopoulos:2007zi}. With the inexact low modes deflation method, we
obtained a big factor of improvement with a $16^3\times32\times8$ lattice on a single node machine. However, it turns out to be very difficult to implement effectively for a highly parallel machine because of the complex structure of the little Dirac operator in the case of domain wall fermions. The Dirac operator for DWF $D_{dwf}$ is not positively definite, so the operator we solve has to be $D_{dwf}^\dagger D_{dwf}$, the resulting little Dirac operator has many hopping terms and it
is very ineffective to calculate its inverse. In comparison, the EigCG algorithm only requires a few linear algebra operations and can easily adapt to massively parallel machine no matter what the operators are. So we used it in our calculation. The disadvantage of the EigCG algorithm compared to L\"uscher's  is the huge requirement of memory. Nevertheless, our current machine has sufficient memory even for the largest lattice we are currently working on, so it is not a serious issue.

\begin{figure}[!tbh]
\begin{center}
\includegraphics[width=0.7\textwidth]{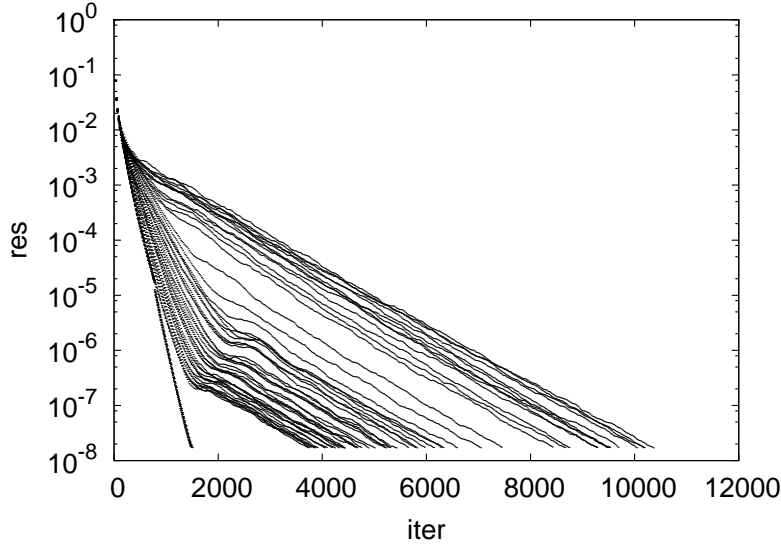}
\end{center}
\caption{Relative residual versus number of iterations using EigCG on a $32^3\times64\times32$ DSDR lattice. From the first 5 propagator solves (60 Dirac solves), the algorithms accumulate more and more low modes. After that, all new solves converge to $10^{-8}$ in roughtly 1500 iterations, by using one restart at $10^{-5}$.}
\label{fig:eigCG_large}
\end{figure}

We follow very closely the original work of EigCG in ~\cite{Stathopoulos:2007zi}. Our goal is to solve $A x=b$ fast for many right hand side vectors b. Here, the operator $A$ we consider is the even odd preconditioned DWF operator $A=D_{dwf}^{pc \dagger}D_{dwf}^{pc}$. The EigCG algorithm works as follows: it accumulates many low modes during each normal CG solve; for each new solve, it projects out the low mode space that the EigCG algorithm already accumulated by an initial solution 
\begin{equation}
x_0 = U(U^\dagger AU)^{-1}U^\dagger b
\label{eq:projection}
\end{equation}
where U projects onto low mode space spanned by the low mode vectors. A typical convergence behavior with EigCG is shown in Figure~\ref{fig:eigCG_large}. The first solve is exactly the same as the normal CG algorithm. The second solve becomes a little bit faster because of the initial projection of the low modes that we already accumulated during the first solve. Gradually, the new solves become faster and faster with more and more low modes available. Finally we will stop accumulating low modes and simply do projections to speed up the calculation.

However, there is a clear turning point (around $res\sim 10^{-6}$) on the convergence curve for the sped-up solves. It dramatically slows down to the normal CG speed at some point. This is because of the inaccuracy of the low modes we obtained from each CG solve. Typically, we try to obtain roughly 16 low modes from each solve and throw away a few with eigenvalue larger than some threshold. It is therefore impossible to get all these 16 low modes very accurately. The
strategy to avoid the slow down with the low accuracy low modes is to do multiple projections by restarting the CG algorithm using the residual of the previous inversion attempt as the new right hand side. Following the initial projection as shown in Eq.\ref{eq:projection}, we do a few more projections in the middle of the solving process. For example, suppose that after n iterations the relative residual reduces to $10^{-5}$, with solution $x_n$ and residual $r_n=b-Ax_n$, we can restart the CG with initial solution $x'_0=x_n+U(U^\dagger AU)^{-1}U^\dagger r_n$ on the equation $A x' =  r_n$. As shown in figure~\ref{fig:eigCG_large}, the relative residual goes straight down once a restart point at $10^{-5}$ is introduced. 

There are two things worth a notice. First, during the low mode accumulation stage, we prefer not to do multiple restart for the solves since it may affect the efficiency of the low modes accumulation. This is the reason that there are turning points in the first 60 convergence curve (except the one goes straight down in 1500 iterations) in figure~\ref{fig:eigCG_large}. Second, if the low modes are extremely inaccurate, we have to do many projections by restarting the CG
algorithm. In the worst case, we may need to do one projection after each CG step. Then we could better incorporate the projection operator in the original operator to perform a so called oblique projection as Luscher's algorithm does~\cite{Luscher:2007se}. On the other hand, for each restart of the CG, we lose all previous information about the direction vectors of the CG algorithm (which is the advantage of the CG to the steepest descent algorithm), so it leads to a decrease of efficiency of the CG algorithm. Therefore, it is better to do fewer restart, only when it is necessary.

We have shown in figure~\ref{fig:eigCG_large} that we could successfully apply EigCG to a $32^3\times 64\times32$ lattice, and gain a factor of 7 speedup. The number of low modes we accumulate, the required memory to achieve this, and the comparison of the number of iterations to the original CG is summarized in table~\ref{tab:EigCG}. Notice that to reduce the memory requirement, we used single precision to store the low modes. This has no negative effect on the 
EigCG algorithm since the low modes we obtained are not very accurate any way. The largest lattice we tested ($32^3\times64\times32$) requires 2 Tbytes memory, the code runs efficiently on 4k BGL nodes, which provides 4 Tbytes memory. 

\begin{table}[!t]
\label{tab:EigCG}
\caption{The speedup from EigCG algorithm on different lattices. $N_{prop}$ stands for the number of propagator solves to get the required number of low modes $N_{low}$. The symbol $*$ means that it is a quenched calculation.}
\begin{center}
\begin{tabular}{ccccccc}
\hline
Lattice & $m_\pi$ & CG & $N_{low}(N_{prop})$ & Total Memory & EigCG & speed up\\
\hline
$16^3\times32\times16$ & 421 MeV & 1840 & 120(1) & 12 GB &  370 & 5.0 \\
$16^3\times32\times16$ & $204^*$ MeV & 3200 & 120(1) & 12 GB &  460 & 7.0 \\
$24^3\times64\times16$ & 330 MeV & 2900 & 400(4) & 272 GB & 530 & 5.5 \\
$32^3\times64\times32$ & 180 MeV & 10400 & 600(5) & 2 TB &1480 & 7.0 \\
\hline
\end{tabular}
\end{center}
\end{table}

\section{Time separated $\pi-\pi$ source}
\begin{figure}[!tbh]
\hskip0.7in
\begin{tabular}{cc}
\includegraphics[width=0.4\textwidth]{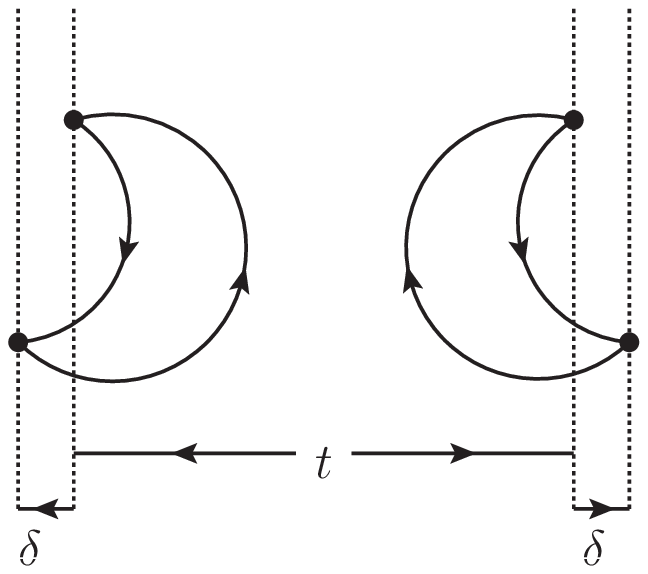} &
\includegraphics[width=0.4\textwidth]{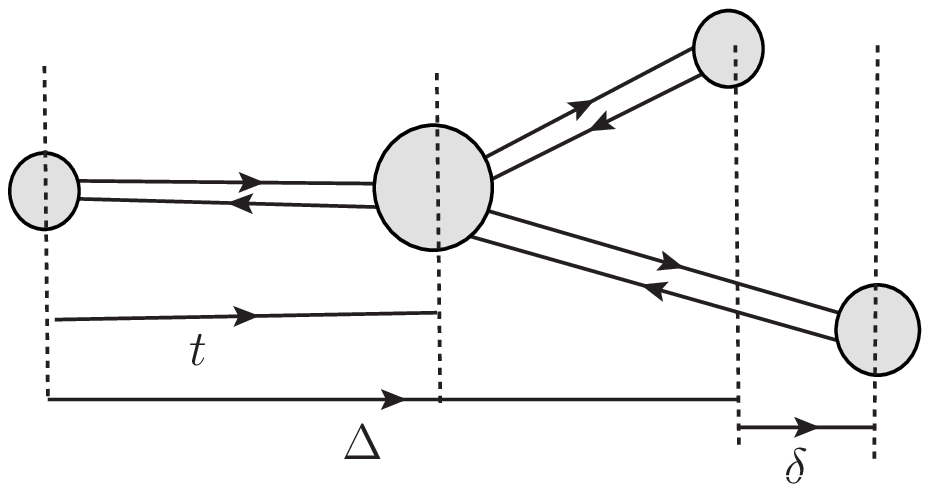}
\end{tabular}
\caption{Separating the two pion sources in the time direction. The left panel shows the setup for the $\pi-\pi$ scattering calculation, and the right panel shows the setup for the $k\rightarrow\pi\pi$ decay calculation.}
\label{fig:sep}
\end{figure}

\begin{figure}[!tbh]
\begin{tabular}{cc}
\includegraphics[width=0.5\textwidth]{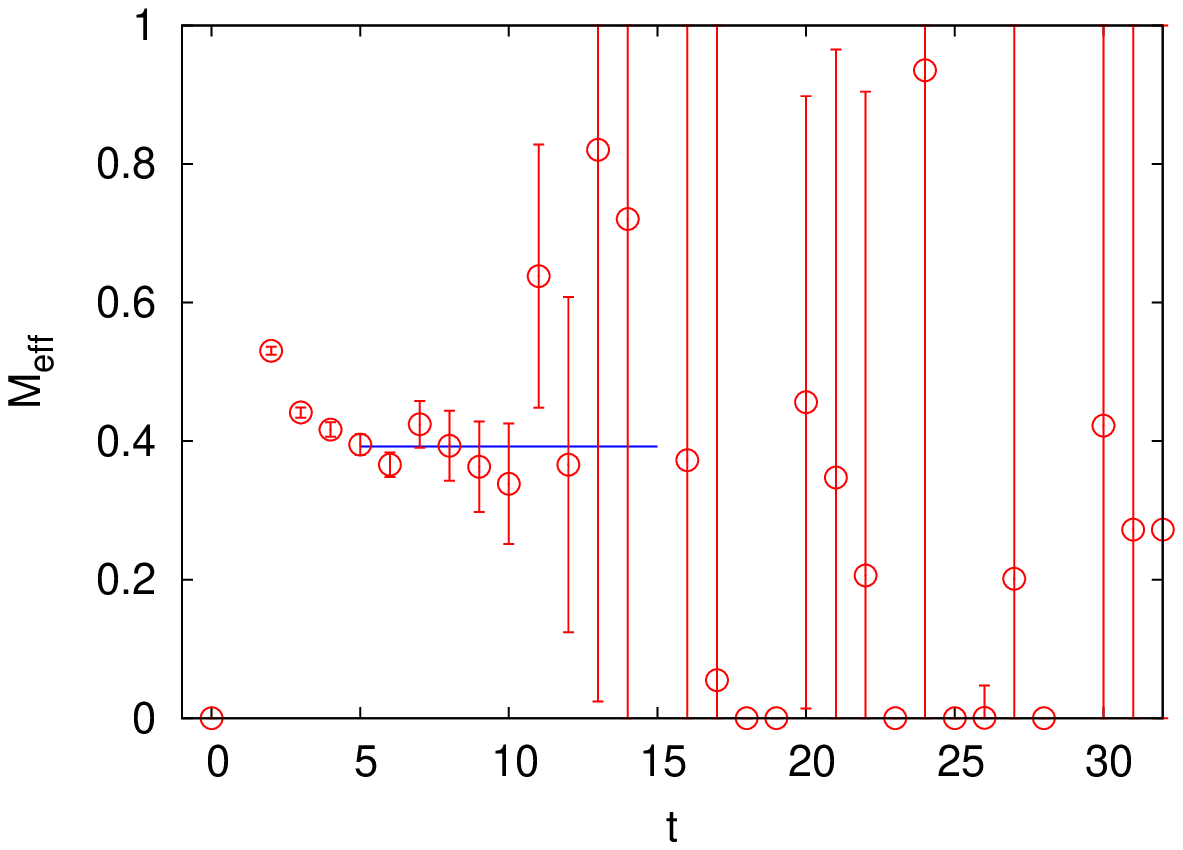} &
\includegraphics[width=0.5\textwidth]{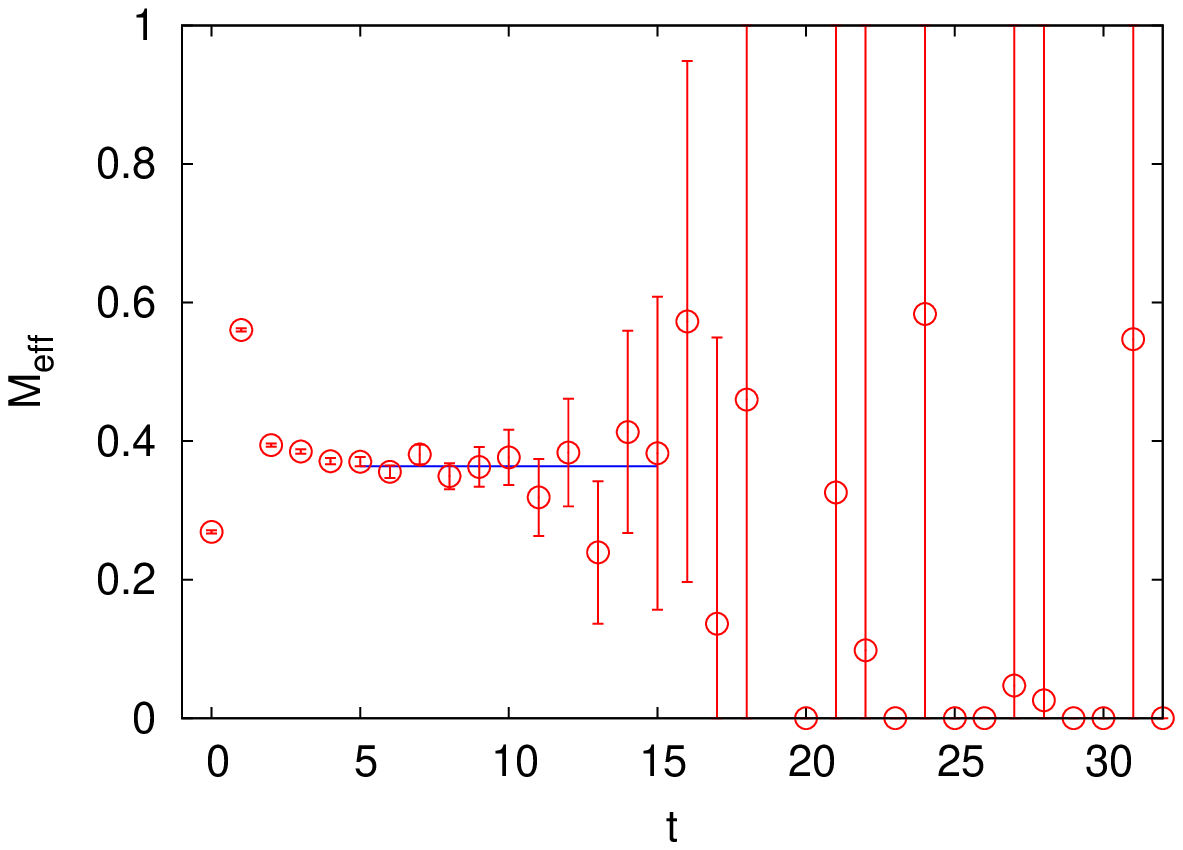}
\end{tabular}
\caption{Effective mass plot for the two pions in the isospin zero channel. The left one uses $\pi-\pi$ separation 0, and the right one uses 4. The energy calculated from these two setups is 0.3922(126) and 0.3639(55) respectively.}
\label{fig:twopion}
\end{figure}

We separate the two pion sources in the time direction by $\delta$ (figure~\ref{fig:sep}) , therefore reducing the correlation between the two pion sources. It can dramatically reduce the vacuum noise from the disconnected graph. For example, the error on the isospin zero $\pi-\pi$ energy is reduced from 0.0126 to 0.0055 by introducing a separation of 4 between the two pions. As shown in figure~\ref{fig:twopion}, the effective mass plateau also begins earlier, even though we still use a fixed fitting range 5 to 15, inclusive. 

\section{$K^0$ to $\pi\pi$ decay amplitudes and conclusion}
Using the techniques we have mentioned, we performed a threshold $m_K=2 m_\pi$,  $K\to\pi\pi$ calculation on a $N_f=2+1$ flavor $24^3\times64\times16$ lattice with DWF,  Iwasaki gauge action, $a^{-1}=1.729(30)$ GeV, and a $330$ MeV pion mass. The EigCG algorithm speeds up the calculation  by a factor of 5, and introducing a separation between the two pion sources by 4 makes the signal much better.

\begin{figure}[!tbh]
\begin{tabular}{cc}
\includegraphics[width=0.5\textwidth]{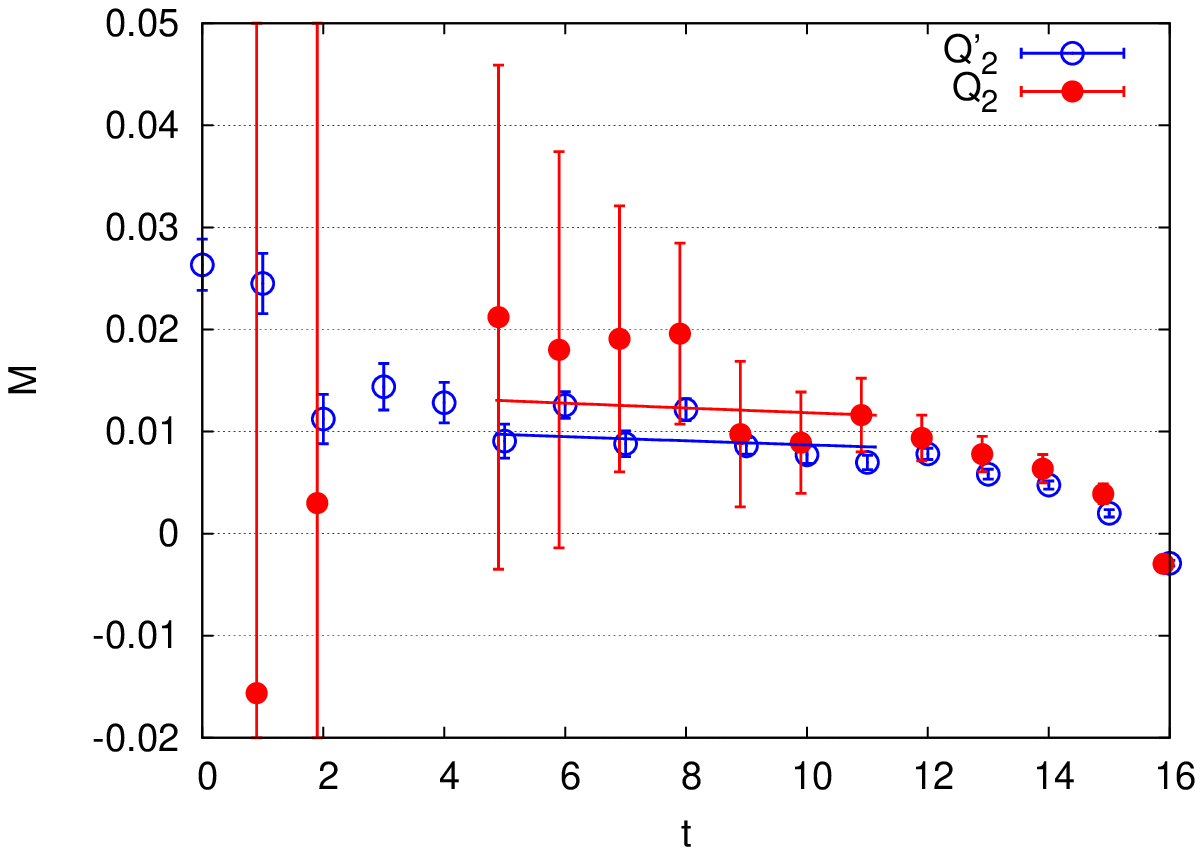} &
\includegraphics[width=0.5\textwidth]{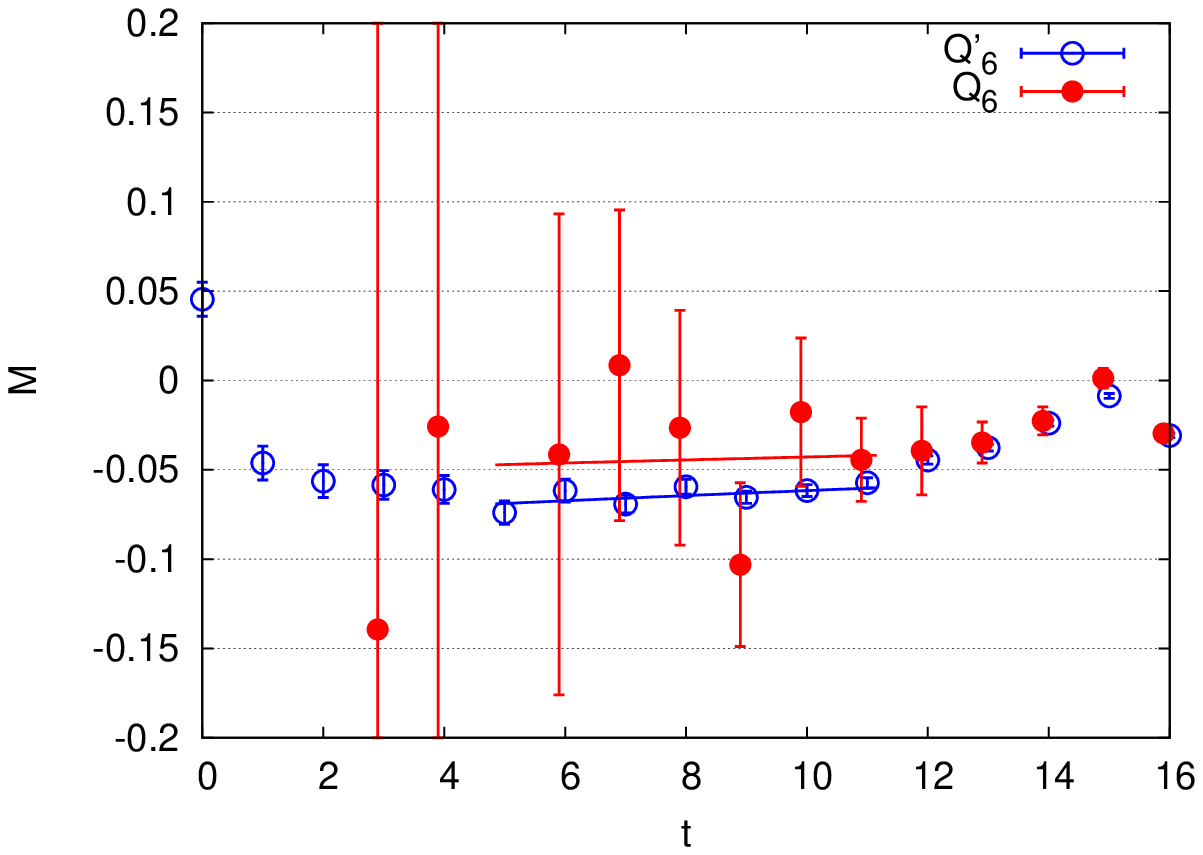} \\
\end{tabular}
\caption{The weak matrix element for $<\pi\pi_{I=0}|Q_2|K^0>$ (left) and $<\pi\pi_{I=0}|Q_6|K^0>$ (right). The x-axis represents the position of the operator relative the the kaon, and y-axis is the amplitude defined in Eq.~\protect\ref{eq:amp}. The $'$ symbol represents the result without the disconnected graph. We used $\Delta = 16$, and $\delta = 4$ here. }
\label{fig:Q2Q6}
\end{figure}

Once we calculate the correlation functions, we do a single parameter fit to find the weak matrix elements,
\begin{equation}
\frac{\left<O_K(0)Q_i(t_{op})O_{\pi\pi}(\Delta,\Delta+\delta)\right>}{N_{\pi\pi} N_K e^{-E_{\pi\pi}\Delta} } = M_i^{1/2,{\rm lat}}  e^{-(m_K-E_{\pi\pi})t}
\label{eq:amp}
\end{equation}
where the kaon energy and $\pi-\pi$ energy are fitted from the kaon and $\pi\pi$ correlation functions. Results for operator $Q_2$ which makes a major contribution to $Re(A_0)$ and the operator $Q_6$ which makes a major contribution to $Im(A_0)$ are shown in figure~\ref{fig:Q2Q6}. A summary of the final results obtained by combing NPR and Wilson coefficients are shown in table~\ref{tab:ktopipi}. This calculation is performed on 138 configurations.

In summary, we performed a full first principle calculation for both $A_2$ and $A_0$ in a 2.7 fm box, with a 660 MeV kaon decaying to two 330 MeV pions. The agreement of the results with and without disconnected graphs indicats that the diconnected graphs may not play a crucial role in this particular decay process. A ratio of 12.0(1.7) for $Re(A_0)$ to $Re(A_2)$ suggests already a dramatic $\Delta I=1/2$ rule effect. The direct CP violation measure
$Re(\epsilon^\prime/\epsilon)$ is calculated to be $2.0(1.7)\times 10^{-3}$ for these unphysical kinematics. In the future, we are going to collect more statistics to resolve a clear signal for $\epsilon'$ and then move to a calculation with physical kinematics.

\begin{table}[!thb]
\label{tab:ktopipi}
\caption{$K^0\rightarrow\pi\pi$ decay amplitudes for a threshold calculation with $m_k\approx 2m_\pi$. The unit for Real part is $\times10^{-8}$ GeV, and Imaginary part is $\times10^{-12}$GeV. The symbol $'$ indicates that the disconnected graphs are ignored. }
\begin{center}
\begin{tabular}{cccccccc}
\hline
$m_\pi$(MeV) & $m_K$(MeV) & Re($A_0$) & Re($A_0'$) & Im($A_0$) & Im($A_0'$) & Re($A_2$) & Im($A_2$) \\
\hline
329.3 & 662.1 & 31.1(4.5) & 27.8(0.8) & -33(15) & -36.3(16) & 2.668(14) & -0.6509(34) \\
\hline
\end{tabular}
\end{center}
\end{table}

{\bf Acknowledgements}
I thank very much all my colleagues in the RBC and UKQCD collaborations for discussions, suggestions, and help. I especially thank my advisor prof. Norman Christ for detailed instructions and discussions. I acknowledge Columbia University, RIKEN, BNL, ANL and the U.S. DOE for providing the facilities on which this work was performed. This work was supported in part by U.S. DOE grant DE-FG02-92ER40699. Finally, I would like to thank the U.S. DOE for support as a DOE Fellow in High Energy Theory.

\bibliography{citations}

\providecommand{\href}[2]{#2}\begingroup\raggedright\begin{thebibliography}{1}

\bibitem{Blum:2011pu}
T.~Blum {\em et~al.}, {\it {$K$ to $\pi\pi$ Decay amplitudes from Lattice
  QCD}},  \href{http://arXiv.org/abs/1106.2714}{{\tt arXiv:1106.2714
  [hep-lat]}}.

\bibitem{Elaine2011}
E.~Goode, {\it {}},  {\em PoS} {\bf LATICE 2011} (2011) 313.

\bibitem{Buchalla:1995vs}
G.~Buchalla, A.~J. Buras and M.~E. Lautenbacher, {\it {Weak decays beyond
  leading logarithms}},  {\em Rev. Mod. Phys.} {\bf 68} (1996) 1125--1144
  [\href{http://arXiv.org/abs/hep-ph/9512380}{{\tt arXiv:hep-ph/9512380}}].

\bibitem{Luscher:2007se}
M.~Luscher, {\it {Local coherence and deflation of the low quark modes in
  lattice QCD}},  {\em JHEP} {\bf 0707} (2007) 081
  [\href{http://arXiv.org/abs/0706.2298}{{\tt arXiv:0706.2298 [hep-lat]}}].

\bibitem{Stathopoulos:2007zi}
A.~Stathopoulos and K.~Orginos, {\it {Computing and deflating eigenvalues while
  solving multiple right hand side linear systems in Quantum Chromodynamics}},
  {\em SIAM J. Sci. Comput.} {\bf 32} (2010) 439--462
  [\href{http://arXiv.org/abs/0707.0131}{{\tt arXiv:0707.0131 [hep-lat]}}].

\end{thebibliography}\endgroup
\bibliographystyle{h-physrev}

\end{document}